\documentclass[aps,nofootinbib,showpacs,twocolumn,floatfix,superscriptaddress]{revtex4}
\usepackage{graphicx}
\usepackage{epsf,amsfonts,amssymb,amsbsy}
\usepackage[mathscr]{eucal}
\begin{document}
\title{Quasiattractor dynamics of $\lambda\phi^4$-inflation}
\author{V.V.Kiselev}
 \affiliation{Russian State Research
Center ``Institute for High
Energy Physics'', 
Pobeda 1, Protvino, Moscow Region, 142281, Russia\\ Fax:
+7-4967-742824}
 \affiliation{Moscow Institute of Physics and
Technology, Institutskii per. 9, Dolgoprudnyi, Moscow Region,
141701, Russia}
\author{S.A.Timofeev}
\affiliation{Moscow Institute of Physics and Technology,
Institutskii per. 9, Dolgoprudnyi, Moscow Region, 141701, Russia}
 \pacs{98.80.-k}
\begin{abstract}
At high e-foldings of expansion, the inflation with the quartic
potential exhibits the parametric attractor governed by the slowly
running Hubble rate. This quasiattractor simplifies the analysis
of predictions for the inhomogeneity generated by the quantum
fluctuations of inflaton. The method reveals the connection of
inflation e-folding with general parameters of preheating regime
in various scenarios and observational data.
\end{abstract}
\maketitle

\section{Introduction}

The homogeneous component of inflationary dynamics in the simplest
case of single field
\cite{i-Guth,i-Linde,i-Albrecht+Steinhardt,i-Linde2} (see review
in \cite{inflation}) is usually described in terms of slow-roll
approximation in the evolution of scalar inflaton $\phi$, when one
neglects both its kinetic energy $\frac{1}{2}\dot\phi^2$ and
acceleration $\ddot\phi$ in the field equations
\begin{equation}\label{intro-1}
\begin{array}{rcl}
  H^2 & \hskip-3pt= & \hskip-3pt \displaystyle
  \frac{8\pi}{3\,m_{\mathtt{Pl}}^2}\left\{\frac{1}{2}\,\dot\phi^2+V(\phi)\right\},\\[4mm]
  \ddot\phi & \hskip-3pt= & \hskip-3pt \displaystyle
  -3H\dot\phi-\frac{\partial V}{\partial \phi},\\[1mm]
\end{array}%
\end{equation}
derived from the Hilbert--Einstein action of
Friedmann--Robertson--Walker metric
\begin{equation}\label{intro-2}
    \mathrm{d}s^2=\mathrm{d}t^2-a^2(t)\,\mathrm{d}\boldsymbol r^2,
\end{equation}
with the scale factor of expansion $a(t)$ ordinary defining the
Hubble rate $H=\dot a/a$ in terms of its derivative with respect
to time $t$ as denoted by over-dot
$\dot\diamond\equiv\mathrm{d}\diamond/\mathrm{d}t$. In
(\ref{intro-1}) the Planck mass is usually introduced through the
Newton constant $G$ by the relation $m_{\mathtt{Pl}}^2=1/G$, while
$V$ is the inflaton potential. The slow-rolling implies
\begin{equation}\label{intro-3}
\begin{array}{rcl}
  H^2 & \hskip-3pt\approx & \hskip-3pt \displaystyle
  \frac{8\pi}{3\,m_{\mathtt{Pl}}^2}\,V,\\[4mm]
  \dot\phi & \hskip-3pt\approx & \hskip-3pt \displaystyle
  -\frac{1}{3H}\,\frac{\partial V}{\partial \phi}.\\[1mm]
\end{array}%
\end{equation}
For the consistency of such the approximation we have to estimate
the ratios
\begin{equation}\label{intro-4}
    \frac{\dot\phi^2}{2V}\approx\frac{1}{3}\,\frac{m_{\mathtt{Pl}}^2}{16\pi}\,
    \left(\frac{\partial\ln V}{\partial \phi}\right)^2\equiv
    \frac{1}{3}\,\epsilon\ll 1,
\end{equation}
and
\begin{equation}\label{intro-5}
    \frac{\ddot\phi}{3H\dot\phi}\approx-\frac{1}{9H^2}\left\{\frac{\partial^2
    V}{\partial\phi^2}-\frac{4\pi}{3m_{\mathtt{Pl}}^2}\,\frac{1}{H^2}\left(
    \frac{\partial V}{\partial \phi}\right)^2\right\}.
\end{equation}
Then, introducing parameter $\eta$ by
\begin{equation}\label{intro-6}
    \eta=\frac{m_{\mathtt{Pl}}^2}{8\pi}\,\frac{1}{V}\,\frac{\partial^2
    V}{\partial \phi^2},
\end{equation}
and using (\ref{intro-3}), we get
\begin{equation}\label{intro-7}
    \frac{\ddot\phi}{3H\dot\phi}\approx-\frac{1}{3}\{\eta-\epsilon\}.
\end{equation}
Therefore, the slow-roll approximation is substantiated at
\begin{equation}\label{intro-8}
    \epsilon\ll 1,\quad |\eta|\ll 1.
\end{equation}
Constraints (\ref{intro-8}) are usually satisfied for positive
power potentials, for instance, at appropriately chosen parameters
and fields $\phi\gtrsim m_{\mathtt{Pl}}$, which guarantee the
inflation of Universe with slowly changing Hubble rate with a huge
total e-folding of scale factor $N_{\mathrm{tot.}}=\ln
a_{\mathrm{end}}/a_{\mathrm{init.}}$, where subscripts stand for
the moments of ending and initiating the inflation,
correspondingly.

Note, that the above speculations have not involved any initial
data on the field evolution. This can be meaningful, if only the
evolution possesses properties of attractors, which allow a stable
dynamical behavior, that soon forgets about a start.

The quasiattractor for the inflationary dynamics with the
quadratic potential $V\sim m^2\phi^2$ was found in
\cite{Mexicans}, which modernized the consideration of dependence
on the initial conditions in the homogeneous case as was done in
pioneering papers of \cite{BGKhZ,PW}. So, the phase space
variables of dynamical system rapidly tend to a stable point,
which position depends on the parameter defined by $H$, while the
parameter itself slowly evolves during the most amount of
e-folding in the inflation and it begins significantly to change
to the end of inflation at a short increment of
$N_{\mathrm{tot.}}$, only. Thus, at high total e-folding of
inflation governed by an appropriate large starting value of
Hubble rate, the inflation due to the quadratic potential is well
described by the quasiattractor behavior. Similar result was
obtained for the quartic potential in \cite{KLS} for the
homogeneous case, and further attempts to generalize the analysis
to damping of initial inhomogeneities were formulated.

The analysis of dynamical system for the homogeneous inflaton and
baryotropic matter was done in \cite{laM-P}, wherein some general
features of evolution were derived for an arbitrary potential of
inflaton, and several explicit examples were studied.

Another aspect of considering the initial conditions concerns for
inhomogeneities perturbing both the inflaton and metric. The
analysis of inhomogeneity in the initial values of scalar field
beyond the perturbation theory as well as relevant references can
be found in \cite{KBranden}, wherein the authors formulated
conditions constraining the parameters of initial state, when the
inflation takes place.

In the present paper we generalize the approach of \cite{Mexicans}
to the case of quartic potential $V\sim \lambda\phi^4$, that
modernizes the consideration in \cite{KLS}. The main goal of our
study is to show the effectiveness and elegance of quasiattractor
method to the analysis of inflation dynamics and its connection
with the preheating regime in the case of quartic potential in
order to draw the conclusion on the consistency with the
observational data versus general conditions of preheating. The
dynamical system relevant to the evolution of homogeneous Universe
is considered in Section \ref{II}. We find the stable
quasiattractor for the quartic potential, too. The inflation
parameters are investigated in Section \ref{III}, wherein they are
confronted with both the experimental data and predictions of
slow-roll approximation. Our results are summarized and discussed
in Conclusion.

\section{Dynamical system\label{II}}

The action of inflaton is of the form
\begin{equation}\label{d-1}
    S=\int \mathrm{d}^4x \sqrt{-g} \left\{\frac{1}{2}\,\partial_\mu\phi \,\partial^\mu\phi -
    V(\phi)\right\},
\end{equation}
with the quartic potential
\begin{equation}\label{d-2}
    V(\phi)=\frac{\lambda}{4}\,\phi^4.
\end{equation}
In the Friedmann--Robertson--Walker metric it leads to the
evolution
\begin{equation}\label{evol}
\begin{array} {rcl}
    \ddot\phi&\hskip-3pt=&\hskip-3pt\displaystyle
    -3H\dot\phi-\lambda\phi^3,\\[3mm]
    \dot H &\hskip-3pt=&\hskip-3pt\displaystyle -\frac{1}{2}\,\kappa^2 \dot\phi^2,
\end{array}
\end{equation}
under the Friedmann equation
\begin{equation}\label{evol-2}
    H^2=\frac{2}{3}\,\kappa^2\left\{\frac{1}{2}\dot \phi^2+\frac{\lambda}{4}\,\phi^4
    \right\},
\end{equation}
at $\kappa^2=8\pi G$.

Following the method of \cite{Mexicans}, let us introduce the
phase space variables
\begin{equation}\label{evol-3}
    x=\frac{\kappa}{\sqrt{6}}\,\frac{\dot \phi}{H}, \qquad
    y=\sqrt[4]{\frac{\lambda}{12}}\frac{\kappa\,\phi}{\sqrt{\kappa H}},
\end{equation}
related due to (\ref{evol-2}) by the constraint
\begin{equation}\label{evol-4}
    x^2+y^4=1.
\end{equation}
The analysis of dynamical system is simplified, if we introduce
the driving parameter
\begin{equation}\label{evol-5}
    z=\frac{\sqrt[4]{3\lambda}}{\sqrt{\kappa H}}.
\end{equation}
Then, denoting the derivative with respect to the amount of
e-folding by prime
$\diamond^\prime\equiv\textrm{d}\diamond/\textrm{d}N$ at $N=\ln
a-\ln a_{\mathrm{init.}}$, we get
\begin{equation}\label{evol-6}
\begin{array}{rcl}
    x^\prime&\hskip-3pt=&\hskip-3pt\displaystyle
    3x^3-3x-2y^3z,\\[3mm]
    y^\prime&\hskip-3pt=&\hskip-3pt\displaystyle
    \frac{3}{2}\,x^2y+xz,\\[3mm]
    z^\prime&\hskip-3pt=&\hskip-3pt\displaystyle
    \frac{3}{2}\,x^2z.
\end{array}
\end{equation}
It is an easy task to confirm, that the system of (\ref{evol-6})
conserves the Friedmann constraint in the form of (\ref{evol-4}).

It is important to stress, that $z$ is quasi-constant, if $x$ is
fixed at a critical point $x_c\ll 1$, hence,
\begin{equation}\label{evol-7}
    z\approx z_{\mathrm{init.}}e^{^{\textstyle{\frac{3}{2}x_c^2N}}}\approx
    z_{\mathrm{init.}}\left(1+\frac{3}{2}\,x_c^2
    N+\ldots\right),
\end{equation}
so that the monotonic growth of $z$ is slow in comparison with the
linear increase of e-folding $N$ due to a small slope proportional
to $x_c^2\to 0$, indeed, and one can put $z\approx
z_{\mathrm{init.}}$ until $\frac{3}{2}\,x_c^2 N\ll 1$, i.e. at
rather large intervals of $N$. We will see later, in fact, that
the stable critical point $x_c^2$ scales as
$x_c^2\approx\beta/(N_{\mathrm{tot.}}-N)$ at
$N_{\mathrm{tot.}}-N\gg 1$ and $\beta=\frac{1}{3}$, so that the
growth takes the form
\begin{equation}\label{evol-7a}
    z\approx z_{\mathrm{init.}} \left(\frac{N_{\mathrm{tot.}}}{N_{\mathrm{tot.}}-N}\right)^{^{
    \textstyle{\frac{3}{2}\beta }}}.
\end{equation}
Therefore, at $N\ll N_{\mathrm{tot.}}$ the increase is quite
linear, but the coefficient of $z_{\mathrm{init.}}$ is actually
suppressed $z_{\mathrm{init.}}\ll 1$ by the conditions of
attractor, as we will consistently see later.

Thus, we can treat the phase-space evolution in the $\{x,y\}$
plane as the autonomous dynamical system with external parameter
$z$, when the driftage of $z$ gives rise to the sub-leading
correction.

\subsection{Critical points}

Solving $x^\prime=y^\prime=0$ at $x\neq0$ and $y\neq0$, we get the
relation\footnote{The critical point $x=y=0$ is off interest,
since it does not satisfy constraint (\ref{evol-4}).}
\begin{equation}\label{crit-1}
    3xy=-2z,
\end{equation}
while
\begin{equation}\label{crit-2}
    f(x^2)\equiv x^6-x^4+\left(\frac{2}{3}\,z\right)^4=0.
\end{equation}
The cubic polynomial $f(u)$ in (\ref{crit-2}) produces two
extremal points, since its derivative is equal to
$$
    \frac{\mathrm{d}f}{\mathrm{d}u}=u(3u-2),
$$
hence, $u_1=0$, $u_2=\frac{2}{3}$. Therefore, $f$ has got positive
roots, if only
$$
    f(u_2)<0,
$$
that yields
\begin{equation}\label{crit-3-0}
    z^4<\frac{3}{4}.
\end{equation}
Constraint (\ref{crit-3-0}) guarantees the existence of critical
points $x_c^2>0$.

Substituting
\begin{equation}\label{crit-3}
    z^4=\frac{3}{4}\sin^2\chi,\quad\mbox{at}\quad
    0<\chi\leqslant\frac{\pi}{2},
\end{equation}
allows us to write the real solutions of (\ref{crit-2}) in the
following form
\begin{equation}\label{crit-4}
    x^2_\ell=\frac{1}{3}\left\{1+2\cos\left(\frac{2}{3}\,\chi+
    \frac{2}{3}\,\pi\ell\right)\right\},\quad \ell=\{0,\pm1\}.
\end{equation}

At $\ell=+1$ quantity $x_+^2\leqslant 0$, which is not a physical
point, at all, while at $\ell=-1$ and $\ell=0$
\begin{equation}\label{crit-5}
    0<x_-^2\leqslant \frac{2}{3},\qquad\frac{2}{3}
    \leqslant x_0^2<1,
\end{equation}
correspondingly. Therefore, the case of $\ell =0$ is covered by
the case of $\ell=-1$, if we expand the interval of $\chi$ to
\begin{equation}\label{crit-6}
    0<\chi<\pi,
\end{equation}
that allows us to put the critical point\footnote{One could change
the convention by setting $-\pi<\chi<0$ and $x_c=x_0$, but we
prefer for (\ref{crit-6}), when the case of interest with
$x_c^2\ll 1$ is spectacularly reached at $\chi\to 0$.} to
$x_c=x_-$.

Due to (\ref{crit-1}), we can write down
$$
    f(x^2)=x^4(x^2+y^4-1),
$$
so that the critical points satisfy the Friedmann constraint of
(\ref{evol-4}).

\subsection{Stability analysis}

At $x=x_c+\bar x$ and $y=y_c+\bar y$, the evolution linear in
$\{\bar x, \bar y\}$ reads off as
\begin{equation}\label{stab-1}
    \left(
\begin{array} {l}
    \bar x ^\prime \\[2mm] \bar y ^\prime
\end{array}
    \right) = \left(
\begin{array}{ccc}
    9x_c^2-3 && -6y_c^2z \\[2mm] -z && \frac{3}{2}x_c^2
\end{array}
\right) \left(
\begin{array}{r}
\bar x \\[2mm] \bar y
\end{array}
\right),
\end{equation}
wherein we have used (\ref{crit-1}).

The linearized Friedmann constraint in (\ref{evol-4}) gives
\begin{equation}\label{fried-1}
    x_c\bar x=-2y_c^3\bar y,
\end{equation}
so that (\ref{stab-1}) is reduced to the single equation
\begin{equation}\label{fired-2}
    \bar y^\prime=\frac{9}{2}\,\left(x_c^2-\frac{2}{3}\right)\bar
    y.
\end{equation}
Therefore, at
\begin{equation}\label{stab-7}
    x_c^2<\frac{2}{3},
\end{equation}
the deviations $\{\bar x,\bar y\}$ will be damped, while at
$x_c\to 0$, the fluctuations will decline as $1/N^3$.

The same fact can be found in a general manner. So, the
eigenvalues of matrix in (\ref{stab-1}) are the following:
\begin{equation}\label{stab-2}
    \lambda_+=6x_c^2,\qquad
    \lambda_-=\frac{9}{2}\left(x_c^2-\frac{2}{3}\right),
\end{equation}
while the corresponding eigenvectors are given by
\begin{equation}\label{stab-3}
    v_+= \left(
    \begin{array}{ccc}
    -\frac{9}{2}\,x_c^2 \\[1mm] z
    \end{array}
    \right),\qquad
    v_-= \left(
    \begin{array}{ccc}
    3y_c^4 \\[1mm] z
    \end{array}
    \right).
\end{equation}
However, the condition of (\ref{fried-1})
can be rewritten as
\begin{equation}\label{fried-3}
    \left(
    \begin{array}{r}
    \bar x \\[2mm] \bar y
    \end{array}
    \right)=\left(
    \begin{array}{ccc}
    3y_c^4 \\[2mm] z
    \end{array}
    \right)\,\frac{\bar y}{z}=v_-\,\frac{\bar y}{z}.
\end{equation}
Therefore, the eigenvector $v_+$ is irrelevant to the
consideration, since it drives out of constraint (\ref{fried-1}),
while $v_-$ is consistent with (\ref{fried-1}), and the critical
point is stable at $\lambda_-<0$, i.e. under (\ref{stab-7}).

Thus, we have established the quasiattractor for the quartic
potential of inflaton. The stability of critical point is
controlled by smallness of $z$, that means large enough value of
initial Hubble rate.

\section{Inflation parameters\label{III}}

Staring the evolution of Universe filled by the inflaton from a
high initial Hubble rate, i.e. at $z_{\mathrm{init.}}\to 0$, but
$H^2_{\mathrm{init.}}<m_{\mathtt{Pl}}^2$, causes a rapid entering
the quasiattractor regime, when the expansion is described by the
slow drift of Hubble rate. It means that a small domain of space
is inflationary increased to a huge size, and the inflaton can be
approximated by an almost homogeneous field versus the spatial
coordinates, while its value and time derivative evolve in
accordance with the drift of critical points.

\subsection{Homogeneous limit}

The quasiattractor is the reason for the cosmological evolution
forgets its primary origin to the leading homogeneous
approximation. Nevertheless, one could consider some properties of
attractor dynamics close to the end of inflation, since it further
enters the properties of observed inhomogeneity.

\subsubsection{Acceleration conditions} The accelerated expansion
takes place at $\ddot a>0$, hence,
$$
    \frac{\ddot a}{a}=\dot H+H^2>0,
$$
that is reduced to
\begin{equation}\label{inf-1}
    \frac{\dot H}{H^2}=-\frac{\kappa^2\dot\phi^2}{2H^2}=-3x^2>-1,
\end{equation}
wherein we have used (\ref{evol}). Therefore, the Universe follows
the inflation at
\begin{equation}\label{inf-2}
    x_c^2<x^2_{c,\mathrm{end}}\equiv\frac{1}{3},
\end{equation}
i.e. in the region, where the quasiattractor is in action, indeed.

The accelerated regime ends at
\begin{equation}\label{inf-3}
\begin{array}{rcl}
    y^4_{c,\mathrm{end}}&\hskip-3pt=&\hskip-3pt\displaystyle
    1-x^2_{c,\mathrm{end}}=\frac{2}{3},\\[3mm]
    z^4_{c,\mathrm{end}}&\hskip-3pt=&\hskip-3pt\displaystyle
    \left(\frac{3}{2}\,x_{c,\mathrm{end}}
    y_{c,\mathrm{end}}\right)^4=\frac{3}{8},
\end{array}
\end{equation}
equivalent to $\chi_{\mathrm{end}}=\frac{\pi}{4}$ and
\begin{equation}\label{inf-3y}
    \phi^2_{\mathrm{end}}=
    \frac{8}{\kappa^2},\quad
    H^2_{\mathrm{end}}=
    \frac{8\lambda}{\kappa^2},\quad
    \dot\phi^2_{\mathrm{end}}=
    \frac{48\lambda}{\kappa^4}.
\end{equation}

The inflation condition of (\ref{inf-2}) is supplemented by
accompaniments
\begin{equation}\label{inf-2x}
    y_c^4>\frac{2}{3},\qquad z^4<\frac{3}{8}.
\end{equation}

Note, that to the end of inflation the field takes the value of
the order of Planck mass independently of coupling constant
$\lambda$, while the corresponding squares of Hubble rate and
kinetic energy are scaled linearly in the coupling constant, and
they are actually suppressed to the appropriate Planck mass powers
at $\lambda\ll 1$.

\subsubsection{Amount of e-folding}

Once the attractor is in action, the total amount of e-folding
during the inflation can be obtained by integration of
$z$-component in the system of (\ref{evol-6}) at $x=x_c$, i.e.
\begin{equation}\label{tot-1}
    N_{\mathrm{tot.}}=\frac{2}{3}
    \int\limits_{z_{\mathrm{init.}}}^{z_{\mathrm{end}}}\frac{\mathrm{d}z}{x^2_c
    z}.
\end{equation}
In the limit of $z_{\mathrm{init.}}\to 0$, we deduce (\ref{tot-1})
by transformation to the integration versus $\chi$, so that
\begin{equation}\label{tot-2}
    N_{\mathrm{tot.}}=
    \int\limits_{{\chi}_{_{\mathrm{init.}}}}^{\pi/4}\frac{\cot\chi\,\mathrm{d}\chi}
    {1+2\cos\left(\frac{2}{3}\,\chi-
    \frac{2}{3}\,\pi\right)},
\end{equation}
approximated by
\begin{equation}\label{tot-3}
    N_{\mathrm{tot.}}^{(0)}\approx\frac{\sqrt{3}}{2}
    \int\limits_{{\chi}_{_{\mathrm{init.}}}}^{\pi/4}\frac{\mathrm{d}\chi}
    {\chi^2},
\end{equation}
to the leading order at ${\chi}_{_{\mathrm{init.}}}\to 0$. The
consequent expansion in (\ref{tot-2}) gives
\begin{equation}\label{tot-4}
    \begin{array}{rcl}
      N_{\mathrm{tot.}}^{(0)} & \hskip-3pt= & \hskip-3pt \displaystyle
      \frac{\sqrt{3}}{2}\left\{\frac{1}{\,{\chi}_{_{\scriptstyle\mathrm{init.}}}}-
      \frac{4}{\pi}\right\},\\[5mm]
      N_{\mathrm{tot.}}^{(1)} & \hskip-3pt= & \hskip-3pt \displaystyle
      \frac{\sqrt{3}}{2}\left\{\frac{1}{\,{\chi}_{_{\scriptstyle\mathrm{init.}}}}-
      \frac{4}{\pi}\right\}+\frac{1}{6}\ln
      \frac{4\,{\chi}_{_{\scriptstyle\mathrm{init.}}}}{\pi}.
    \end{array}
\end{equation}
The inversion of (\ref{tot-4}) results, for instance, in
\begin{equation}\label{tot-5}
    {\chi}_{_{\scriptstyle\mathrm{init.}}}^{(0)}=\frac{\pi\sqrt{3}}{2\pi
    N_{\mathrm{tot.}}^{(0)}+4\sqrt{3}}.
\end{equation}

To the leading order at $N_{\mathrm{tot.}}\gg 1$ we deduce the
initial data
\begin{equation}\label{tot-6}
    \begin{array}{rcl}
      z^2_{\mathrm{init.}} & \hskip-3pt\approx & \hskip-3pt\displaystyle
      \frac{3}{4}\,\frac{1}{N_{\mathrm{tot.}}},\qquad
      x^2_{c,\mathrm{init.}}\approx
      \frac{1}{3}\,\frac{1}{N_{\mathrm{tot.}}},
      \\[3mm]
    \end{array}
\end{equation}
while $y^4_{c,\mathrm{init.}}=1-x^2_{c,\mathrm{init.}}$. Note,
that the initial data for the field and its velocity $\{y, x\}$
mean their values \textit{just after entering the quasiattractor},
while the actual data are irrelevant, since the quasiattractor
rapidly adjust the field and its velocity to (\ref{tot-6}) in
accordance to the initial Hubble rate, i.e. $z_{\mathrm{init.}}$.

\subsection{Inhomogeneity}

Quantum fluctuations of inflaton near its homogeneous classical
value result in the primordial spatial perturbations of energy
density, involving the scalar and tensor components of spectrum
versus the wave vector $k$ at the moment, when the fluctuation
comes back from the horizon, i.e. at $k=a(t)\,H$ as following:
\begin{equation}\label{inh-1}
    \begin{array}{rcl}
      \mathcal{P}_{\mathrm{S}}(k) & \hskip-3pt= & \hskip-3pt\displaystyle
      \left(\frac{H}{2\pi}\right)^2\left(\frac{H}{\dot\phi}\right)^2,\\[5mm]
      \mathcal{P}_{\mathrm{T}}(k) & \hskip-3pt= & \hskip-3pt\displaystyle
      8\kappa^2\left(\frac{H}{2\pi}\right)^2.
    \end{array}
\end{equation}
The spectra can be accurately evaluated in terms of quasiattractor
dynamics by
\begin{equation}\label{inh-2}
    \begin{array}{rcl}
      \mathcal{P}_{\mathrm{S}}(k) & \hskip-3pt= & \hskip-3pt\displaystyle
      \frac{\lambda}{8\pi^2}\,\frac{1}{z^4 x_\mathrm{c}^2},\\[5mm]
      \mathcal{P}_{\mathrm{T}}(k) & \hskip-3pt= & \hskip-3pt\displaystyle
      \frac{6\lambda}{\pi^2}\,\frac{1}{z^4},
    \end{array}
\end{equation}
where $z$ corresponds to the Hubble rate $H$, and it is expressed
by amount of e-folding $N$ between the moments of horizon exit and
inflation end, as that can be found in the right analog of
(\ref{tot-6}) in the leading order at $N\gg 1$. So, we redefine
$N=\ln a_{\mathrm{end}}-\ln a$. Then,
\begin{equation}\label{inh-3}
    \begin{array}{rcl}
      \mathcal{P}_{\mathrm{S}}(k) & \hskip-3pt\approx & \hskip-3pt\displaystyle
      \frac{2\lambda}{3\pi^2}\,
      {N^3},\\[5mm]
      \mathcal{P}_{\mathrm{T}}(k) & \hskip-3pt\approx & \hskip-3pt\displaystyle
      \frac{32\lambda}{3\pi^2}\,
      {N^2},
    \end{array}
\end{equation}
which are consistent with the slow-roll approximation for the
inflation with the quartic potential. The sub-leading corrections
are given by the substitution following from (\ref{tot-5}),
\begin{equation}\label{inh-4}
    \frac{1}{N}\mapsto \frac{2\pi}{2\pi N+4\sqrt{3}}.
\end{equation}
The ratio
\begin{equation}\label{inh-5}
    r=\frac{\mathcal{P}_{\mathrm{T}}}{\mathcal{P}_{\mathrm{S}}}=48\,x_{\mathrm{c}}^2
    \ll 1,
\end{equation}
determines the relative contribution of tensor spectrum. In the
leading order
\begin{equation}\label{inh-6}
    r\approx \frac{16}{N},
\end{equation}
in accordance with the slow-roll approximation, again.

The spectral index is defined by
\begin{equation}\label{inh-7}
    n_{\mathrm{S}}-1\equiv \frac{\mathrm{d}\ln
    \mathcal{P}_{\mathrm{S}}}{\mathrm{d}\ln k}.
\end{equation}
It can be calculated under the condition\footnote{Note, that
e-folding counts backward, from the end of inflation to the moment
$t$, hence, the derivatives in (\ref{evol-6}) change sign.}
\begin{equation}\label{inh-7-1}
    \ln\frac{k}{k_{\mathrm{end}}}=-N-2\ln\frac{z}{z_{\mathrm{end}}},
\end{equation}
so that
\begin{equation}\label{inh-8}
    n_{\mathrm{S}}-1=-6
    x_{\mathrm{c}}^2\,\frac{3-4x_{\mathrm{c}}^2}
    {(1-3x_{\mathrm{c}}^2)(2-3x_{\mathrm{c}}^2)},
\end{equation}
that is reduced to the slow-roll result in the leading
approximation
\begin{equation}\label{inh-8sr}
    n_{\mathrm{S}}-1\approx-9x_{\mathrm{c}}^2\approx-\frac{3}{N}.
\end{equation}
To the same approximation
\begin{equation}\label{inh-8b}
    n_{\mathrm{S}}-1\approx-\frac{3}{16}\,r.
\end{equation}

The running of spectral index is given by
\begin{equation}\label{inh-9}
    \alpha=\frac{\mathrm{d}n_{\mathrm{S}}}{\mathrm{d}\ln k}=
    -36x_{\mathrm{c}}^4\frac{(6-16x_{\mathrm{c}}^2+9x_{\mathrm{c}}^4)
    (1-x^2_{\mathrm{c}})}{(2-3x_{\mathrm{c}}^2)^3(1-3x_{\mathrm{c}}^2)^3},
\end{equation}
approximately equal to
\begin{equation}\label{inh-9a}
    \alpha\approx -27x_{\mathrm{c}}^4\approx-\frac{3}{N^2}.
\end{equation}
As it was noted in \cite{Mexicans}, the evaluation of $\alpha$ in
the framework of slow-roll approximation is quite complicated,
while the quasiattractor dynamics allows us to get it
straightforwardly by means of taking the derivative. Though one
could use (\ref{inh-7-1}) to the leading order, when
$\mathrm{d}k/\mathrm{d}N\approx -1$, that straightforwardly yields
the spectral index as well as its slope derived above.

Thus, the quasiattractor behavior provides us with the direct way
of calculating the properties of primary inhomogeneity in terms of
two parameters: the coupling constant $\lambda$ and number of
e-folding $N$ relative to the observational scale $k$.

\subsubsection{Constraining a maximal $N$}

Following the method elaborated in \cite{LL}, let us restrict the
scale $k=a H$ relative to the observation of inhomogeneity, as it
comes from the both known and suggested history of Universe
evolution by considering the ratio
$$
    \frac{k}{a_0H_0}=\frac{a H}{a_0 H_0},
$$
where $H_0$ is the present-date Hubble rate 
\begin{equation}\label{H0}
    H_0=1.75\cdot 10^{-61}\,h\,m_{\mathtt{Pl}},\qquad h\simeq 0.7,
\end{equation}
while $a_0$ is the present-date scale factor, that can be set to
$a_0\equiv 1$ by definition of length unit. We decompose the
evolution in several consequent stages: the inflation, reheating,
radiation dominance up to the epoch of equality with
nonrelativistic matter, matter dominance, that give
\begin{equation}\label{k1}
\begin{array}{l}
    \displaystyle
    \frac{a}{a_0} =
    \frac{a}{a_{\mathrm{end}}}\,\frac{a_{\mathrm{end}}}{a_{\mathrm{reh.}}}\,
    \frac{a_{\mathrm{reh.}}}{a_{\mathrm{eq.}}}\,\frac{a_{\mathrm{eq.}}}{a_0},\\[5mm]
    \displaystyle
    \frac{H}{H_0} =
    \frac{H}{H_{\mathrm{end}}}\,\frac{H_{\mathrm{end}}}
    {H_{\mathrm{eq.}}}\,\frac{H_{\mathrm{eq.}}}{H_0}.\\[3mm]
\end{array}
\end{equation}
The data yield
\begin{equation}\label{k2}
    H_{\mathrm{eq.}}\simeq 5.25\cdot
    10^6\,h^3\Omega^2_{\mathrm{M}}H_0,\quad
    \frac{a_{\mathrm{eq.}}H_{\mathrm{eq.}}}{a_0H_0}\simeq 219\,h\,\Omega_{\mathrm{M}},
\end{equation}
where $\Omega_{\mathrm{M}}$ is the fraction of matter energy. The
quasiattractor gives
\begin{equation}\label{k3}
    \frac{H^2}{H^2_{\mathrm{end}}}=\frac{z^4_{\mathrm{end}}}{z^4}=\frac{2}{3}\,N^2.
\end{equation}
The scale ratios are related with the ratios of Hubble rates in
two subsequent stages $``b"\to ``c"$ by means of
\begin{equation}\label{k4}
    H_b^2=H_c^2\,\left(\frac{a_c}{a_b}\right)^{3(1+w_c)},
\end{equation}
where $w_c$ is the state parameter at the stage $``c"$, determined
by the ratio of pressure to the energy density. For the radiation
stage we set $w_{\mathrm{eq.}}=\frac{1}{3}$.

Then, we arrive to
\begin{equation}\label{k5}
    \begin{array}{rcl}
      N & \hskip-3pt= & \hskip-3pt\displaystyle
      -\ln\frac{k}{a_0H_0}+\ln 219\, \Omega_{\mathrm{M}}h+\frac{1}{2}\ln\frac{2}{3}+\ln N
      \\[5mm]
      & \hskip-3pt+ & \hskip-3pt\displaystyle
      \left(\frac{1}{2}-\frac{1}{3(1+w_{\mathrm{reh.}})}\right)
      \ln\frac{H^2_{\mathrm{end}}}{m^2_{\mathtt{Pl}}}
      \\[5mm]
      & \hskip-3pt- & \hskip-3pt\displaystyle
      \left(\frac{1}{4}-\frac{1}{3(1+w_{\mathrm{reh.}})}\right)
      \ln\frac{H^2_{\mathrm{reh.}}}{m^2_{\mathtt{Pl}}}-\frac{1}{4}
      \ln\frac{H^2_{\mathrm{eq.}}}{m^2_{\mathtt{Pl}}}.
    \end{array}
\end{equation}
Let us parameterize the Hubble rate at the end of reheating by a
scale $\mu_{\mathrm{reh.}}$, so that
\begin{equation}\label{k6}
    H^2_{\mathrm{reh.}}=\frac{8\pi}{3\,m^2_{\mathtt{Pl}}}\,
    \,\mu^4_{\mathrm{reh.}},
\end{equation}
while
\begin{equation}\label{k7}
    \frac{H^2_{\mathrm{end}}}{m^2_{\mathtt{Pl}}}=\frac{3\pi}{2N^3}\,
    \mathcal{P}_{\mathrm{S}}.
\end{equation}
Therefore,
\begin{equation}\label{k5x}
    \begin{array}{rcl}
      N & \hskip-3pt= & \hskip-3pt\displaystyle
      -\ln\frac{k}{a_0H_0}+\ln 219\, \Omega_{\mathrm{M}}h+\frac{1}{2}\ln\frac{2}{3}
      \\[5mm]
      & \hskip-3pt- & \hskip-3pt\displaystyle
      \left(\frac{1}{2}-\frac{1}{1+w_{\mathrm{reh.}}}\right)\ln N
      -\frac{1}{4}
      \ln\frac{H^2_{\mathrm{eq.}}}{m^2_{\mathtt{Pl}}}
      \\[5mm]
      & \hskip-3pt+ & \hskip-3pt\displaystyle
      \left(\frac{1}{2}-\frac{1}{3(1+w_{\mathrm{reh.}})}\right)
      \ln\frac{3\pi}{2}\,\mathcal{P}_{\mathrm{S}}
      \\[5mm]
      & \hskip-3pt- & \hskip-3pt\displaystyle
      \left(\frac{1}{4}-\frac{1}{3(1+w_{\mathrm{reh.}})}\right)
      \ln\frac{8\pi}{3}\,
      \frac{\mu^4_{\mathrm{reh.}}}{m^4_{\mathtt{Pl}}}.
    \end{array}
\end{equation}
Note, that (\ref{k5x}) is independent of current matter density
$\Omega_{\mathrm{M}}$.

Astronomical observations are naturally restricted by the
distances less than the event horizon given by the inverse Hubble
rate of present day $H_0$. Therefore, the maximal value of
e-folding $N_0$ is reached at the wave number $k\mapsto
k_0=a_0H_0$, when (\ref{k5x}) contains two parameters determined
by the mechanism of reheating. In addition, experimental data give
the spectral density \cite{WMAP}
\begin{equation}\label{exp}
    \mathcal{P}_{\mathrm{S}}\simeq 2.5\cdot 10^{-9}
\end{equation}
at the scale $k\simeq 0.05-0.002$ Mpc$^{-1}$ greater than $k_0$,
nevertheless, this value of $\mathcal{P}_{\mathrm{S}}$ rather
slowly evolves with $k$.

\begin{figure}[th]
\setlength{\unitlength}{1.1mm}
  \begin{center}
  \begin{picture}(75,46)
  \put(0,0){\includegraphics[width=70\unitlength]{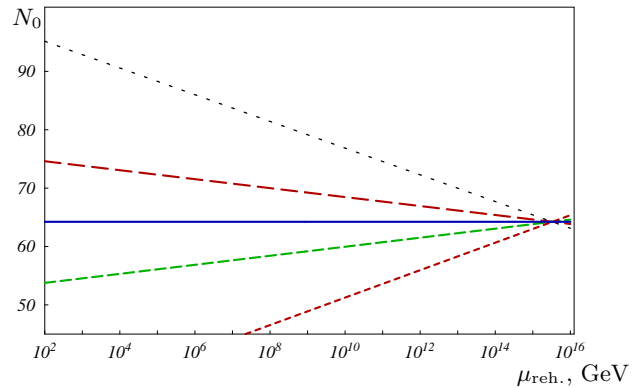}}
  \put(0,41){$N_0$}
  \put(61,-2){$\mu_{\mathrm{reh.}}$, GeV}
  \end{picture}
  \end{center}
  \caption{The maximal number of e-folding $N_0$ versus the reheating scale
  in different regimes of reheating state parameter $w_{\mathrm{reh.}}=1$
  (long-dashed line), $w_{\mathrm{reh.}}=\frac{1}{3}$ (solid line),
  $w_{\mathrm{reh.}}=0$ (dashed line), $w_{\mathrm{reh.}}\to \infty$ (dotted line), and
  $w_{\mathrm{reh.}}=-\frac{1}{3}$ (short-dashed line).}\label{max}
\end{figure}

The dependence of $N_0$ versus the scale of reheating
$\mu_{\mathrm{reh.}}$ and state parameter $w_{\mathrm{reh.}}$ is
shown in Fig. \ref{max} for generic cases of stiff matter $w=1$,
radiation $w=\frac{1}{3}$ and dust $w=0$. It is spectacular, that,
first, the dependence versus the scale is well approximated by
logarithmic law, second, all these reheating regimes converge to a
single point at $\bar\mu_{\mathrm{reh.}}\simeq 0.33\cdot 10^{16}$
GeV, and third, the radiation regime gives the maximal amount of
e-folding\footnote{The value of $N_\mathrm{0,rad.}$ is obtained by
numerical solving (\ref{k5x}) at $w=\frac{1}{3}$.} equal to
$N_\mathrm{0,rad.}\approx 64$ independent of the scale
characterizing the reheating stage. The reheating scale
$\bar\mu_{\mathrm{reh.}}$ for the intersection of curves with
different values of state parameter $w_{\mathrm{reh.}}$ in Fig.
\ref{max} is given by the condition of nullifying the coefficient
in front of factor $1/(1+w_{\mathrm{reh.}})$ in (\ref{k5x}), so
that
\begin{equation}\label{intersection}
    \bar\mu_{\mathrm{reh.}}^4=\frac{9}{16}\,
    \frac{\mathcal{P}_\mathrm{S}}{N_\mathrm{0,rad.}^3}\,
    m^4_\mathtt{Pl}.
\end{equation}

Our numerical calculations exhibit the oscillations of the field
around the minimum of quartic potential with
$w_{\mathrm{reh.}}\approx 1$. However, this value can be modified
by the both corrections to the potential near the minimum and
inflaton interaction with ordinary fields.

We have also shown two marginal cases. The first case corresponds
to $w_{\mathrm{reh.}}=-\frac{1}{3}$ and it could appear, if we
suggest the inflaton oscillations near the potential minimum
neglecting the gravitational damping during the period. Then, the
virial theorem gives averages of kinetic and potential energies as
$\langle x^2\rangle =\frac{1}{2}\langle y^4\rangle$, that
reproduce the mentioned state parameter.

The second case corresponds to a quite instantaneous falling of
Hubble rate to its reheating scale, so that $w_{\mathrm{reh.}}\to
\infty$. Such the regime could happen due to a specific tachyonic
preheating \cite{tach-preh} with a rapid decay of classical
inflaton to quanta, for instance. However, the jump of Hubble rate
is the indication of energy jump, that is unrealistic, at all.
Nevertheless, the situation with $w>1$ takes place, if the
potential minimum is shifted to a negative value
$V_{\mathrm{min}}
0<0$, that gives
$$
    w=\frac{\frac{1}{2}\,\dot\phi^2-V}{\frac{1}{2}\,\dot\phi^2+V}\mapsto
    \frac{\frac{1}{2}\,\dot\phi^2-V_{\mathrm{min}}}{\frac{1}{2}\,
    \dot\phi^2+V_{\mathrm{min}}}>1.
$$
Therefore, the kinetic energy and Hubble rate will rapidly fall
down during short increment of e-folding, while the state
parameter will follow $w\gg 1$, until the potential will become to
grow. Then, we suggest the existence of second local minimum of
$V$ at $V(0)=0$ giving the flat vacuum. The decay of flat vacuum
to the Anti-de Sitter point $V_{\mathrm{min}}<0$ is preserved by
the gravitational effects \cite{CdL,Weinberg}, so that the flat
vacuum can be stable. The barrier between two minima of potential
suggests the presence of negative second derivative of potential
with respect to the field, that can switch on the preheating
mechanism at the scale of potential barrier denoted by
$\mu_{\mathrm{reh.}}^4$, which could be much less than the energy
density at the end of inflation. Thus, in Fig. \ref{max} the
region between the dotted and long-dashed lines can be actual in
the extended version of quartic inflation at
$|V_{\mathrm{min}}|\ll m_{\mathtt{Pl}}^4$. Though, we cannot point
to any verified realistic model of such the scenario to the
moment.

The observational data deal with the wave number $k$ about two
orders of magnitude less than $k_0$, that diminish the
corresponding number of e-folding by the value about $\delta
N\simeq 4-5$.

\subsubsection{Comparing with data}

\begin{figure}[th]
\setlength{\unitlength}{1.1mm}
  \begin{center}
  \begin{picture}(75,75)
  \put(0,0){\includegraphics[width=70\unitlength]{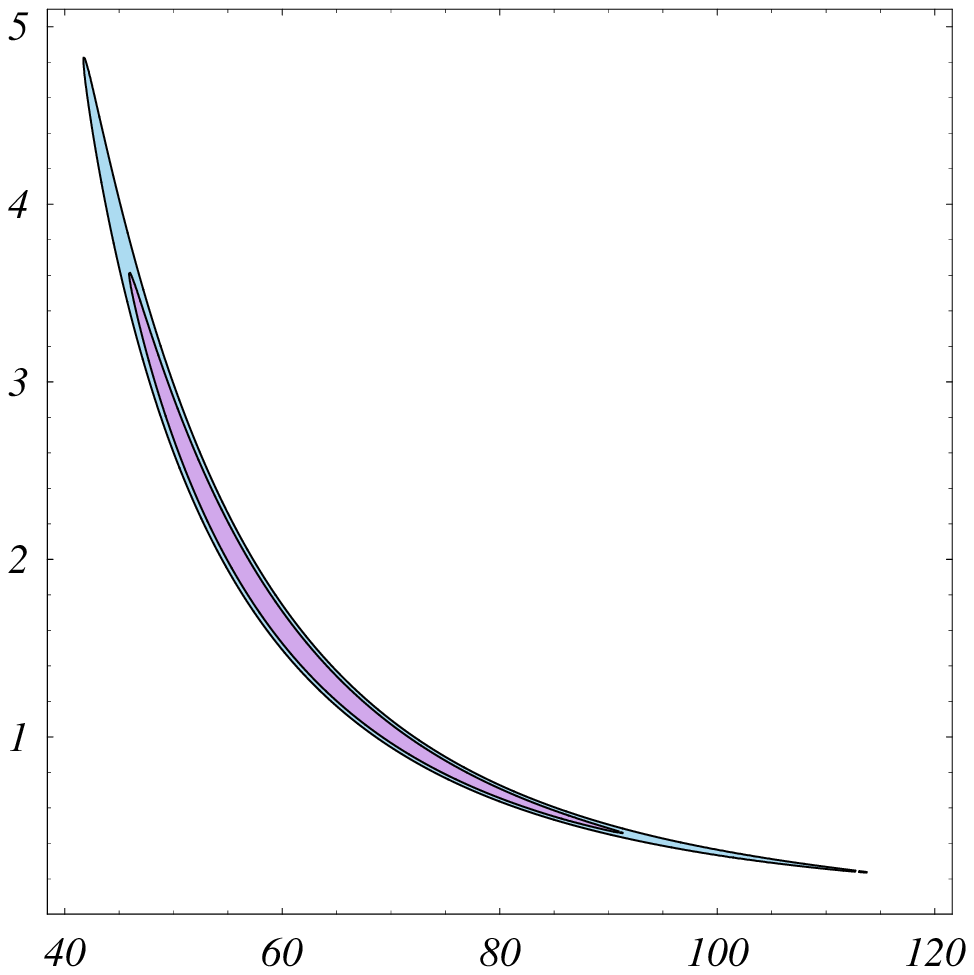}}
  \put(0,72){$\lambda\cdot 10^{13}$}
  \put(61,-2){$N$}
  \multiput(33,4.7)(1,0){35}{\line(0,1){64.3}}
  \end{picture}
  \end{center}
  \caption{The experimentally admissible region
  in the plain of $(N,\lambda)$ with $1\sigma$ and $2\sigma$
  contours as follows from (\ref{dat1}),
  (\ref{dat2}). The shaded region is allowed after taking into account
  the correlation of spectral index $n_{\mathrm{S}}$
  with the fraction of tensor perturbations in the energy density
  $r$ at $2\sigma$ level.}\label{lam-n}
\end{figure}

For the flat Universe with cosmological constant, recent WMAP data
\cite{WMAP} at $k=0.002$ Mpc$^{-1}$ gave quite precise values of
\begin{equation}\label{dat1}
    n_{\mathrm{S}}=0.951^{+0.015}_{-0.019},\qquad
    \mathcal{P}_{\mathrm{S}}=2.36^{+0.12}_{-0.16}\cdot
    10^{-9},
\end{equation}
while the slope of $n_{\mathrm{S}}$ is not so restrictive
\begin{equation}\label{dat2}
    \alpha=-0.102^{+0.050}_{-0.043}.
\end{equation}
Then, we extract the amount of appropriate e-folding
\begin{equation}\label{dat3}
    N=61^{+30}_{-15},
\end{equation}
that yields the coupling constant equal to
\begin{equation}\label{dat4}
    \lambda=1.6^{+2.0}_{-1.2}\cdot 10^{-13},
\end{equation}
which follows from Fig. \ref{lam-n}.

Note, that the inflation predicts the significant suppression of
slope $\alpha$ with respect to the central value in (\ref{dat2}).
Therefore, an improvement of accuracy for the measurement of slope
could serve for discriminating the inflation models.

\begin{figure}[th]
\setlength{\unitlength}{1.1mm}
  \begin{center}
  \begin{picture}(75,45 )
  \put(0,0){\includegraphics[width=70\unitlength]{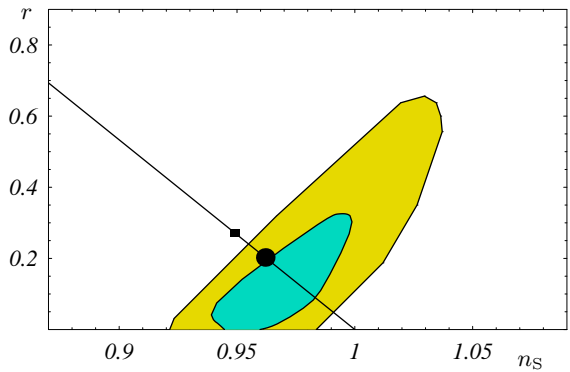}}
  \put(3,42){$r$}
  \put(63,-0){$n_{\mathrm{S}}$}
  \end{picture}
  \end{center}
  \caption{The WMAP data on the correlations between the
spectral index of scalar perturbations $n_{\mathrm{S}}$ and
fraction of tensor perturbations in the energy density $r$, as
represented by $1\sigma$ and $2\sigma$ contours. The line gives
the predictions of quartic inflation versus large amount of
e-folding $N$. The dot stands at $N=79$, while the rectangle does
at $N=59$.}\label{corr}
\end{figure}

Further, the WMAP data gave a strong correlations between the
spectral index of scalar perturbations $n_{\mathrm{S}}$ and
fraction of tensor perturbations in the energy density $r$, that
is shown in Fig. \ref{corr}. The correlations restricts the
regions of $n_{\mathrm{S}}$ admissible for the quartic-potential
inflation, as shown in Fig. \ref{lam-n} by the shaded domain.

Thus, the $\lambda\phi^4$-inflations could be marginally
consistent with observations, if one suggest the extended version
of preheating at low scales about $\mu_{\mathrm{reh.}}\sim 10^9$
GeV with the preliminary passing the region with negative values
of potential as we can conclude from Figs. \ref{max}--\ref{corr}.
An example of model for such the potential will be considered
elsewhere \cite{prepare}. However, to the moment there is no a
safe realistic model for the preheating scenario consistent with
the observational data in the quartic inflation.

\section{Conclusion}

We have shown that the inflationary dynamics with the quartic
potential obeys the parametric quasiattractor governed by the
Hubble rate slowly evolving with e-folding of expansion. The
condition of attractor stability is preserved by the condition of
accelerated expansion.

The quasiattractor allows us to express the inflationary
parameters in terms of coupling constant $\lambda$ and amount of
e-folding $N$ in consistence with the slow-roll approximation.
Sub-leading terms to the approximation are also on hands.

For the case of quartic potential, we have re-analyzed the
possible maximal amount of e-folding $N_0$ corresponding to the
scale of astronomical observations measuring the inhomogeneities
generated by the quantum perturbations of inflaton just before the
end of inflation. It is spectacular that at the reheating scale
$\mu_{\mathrm{reh.}}\sim 0.3\cdot 10^{16}$ GeV, the value of
$N_0\approx 64$ is independent of the particular mechanism of
reheating parameterized by the state parameter
$w_{\mathrm{reh.}}$. At the low-scale reheating the maximal $N_0$
could be increased due to the essential modification of potential
near the origin. For instance, the field could pass the region of
negative potential with the further relaxation in the flat minimum
at $V=0$ after overcoming the barrier producing the tachyonic
preheating.

Then, the standard quartic inflation with realistic parameters of
preheating regime is inconsistent with the observations of
matter-density fluctuations, its spectral index of scalar
perturbations and fraction of tensor fluctuations. However, there
is the marginal case at the coupling constant $\lambda\sim 6\cdot
10^{-14}$ and amount of e-folding $N\sim 80$ in the modified
preheating at low-scales with the negative valley of potential
described above, so that the quartic potential of inflaton at high
fields with the appropriate modification near the origin is
generally still not excluded, but a realistic model with such the
scenario is not known.

This work is partially supported by the Russian Foundation for
Basic Research, grant 04-02-17530.

\end{document}